\documentclass[a4paper]{jpconf}
\usepackage{graphicx}
\def\gvs{g_{\sss{V}}}
\def\gas{g_{\sss{A}}}
\newcommand{\Mass}{\mathrm{M}}

\def\br{\begin{eqnarray}}
\def\er{\end{eqnarray}}
\def\etal{{\it et al. }}
\def\etc{ {\it etc}}
\def\sq{\sqrt{2}}
\def\sss{\scriptscriptstyle}
\def\rb {{\bf r}}
\def\kb {{\bf k}}
\def\lb {{\bf l}}
\def\ga{\overline{g}_{\mbox{\tiny A}}}
\def\gv{\overline{g}_{\mbox{\tiny V}}}

\def\gpa{\overline{g}_{\mbox{\tiny P1}}}
\def\gpb{\overline{g}_{\mbox{\tiny P2}}}
\def\gw{\overline{g}_{\mbox{\tiny W}}}

\def\gA{g_{\mbox{\tiny A}}}

\def\be{\begin{equation}}
\def\ee{\end{equation}}
\def\br{\begin{eqnarray}}
\def\er{\end{eqnarray}}
\def\rf#1{{(\ref{#1})}}
\def\ket#1{|#1 \rangle}
\def\bra#1{\langle #1|}
\def\Ket#1{||#1 \rangle}
\def\Bra#1{\langle #1||}
\def\kb {{\bf k }}
\def\lb{ {\bf l}}
\def\pb {{\bf p}}
\def\qb {{\bf q}}
\def\rb {{\bf r}}
\def\Ob{ {\bf O}}
\def\Jb{ {\bf J}}
\def\mbs{\mbox{\boldmath$\sigma$}}
\def\mbn{\mbox{\boldmath$\nabla$}}
\def\kr {{\bf k}\cdot{\bf r}}

\def\a {{\alpha}}

\def\s {{\sigma}}

\def\w {{\omega}}
\def\k {{\kappa}}

\def\nn{\nonumber }
\def\go{\rightarrow  }
\def\M {{{\cal M}}}
\def\T {{{\cal T}}}
\def\L {{{\cal L}}}
\def\ie{{\em i.e., }}
\begin{document}
\title{Neutrino and antineutrino cross sections in $^{12}$C}
\author{A. R. Samana$^{1}$, F. Krmpoti\'c$^2$, N. Paar$^{3}$, and C. A. Bertulani$^{4}$}
\address{$^1$ Departamento de Cs. Exactas e Tecnol\'ogicas, UESC, Brasil}
\address{$^2$ Instituto de F\'isica La Plata, Universidad Nacional de La Plata, La Plata, Argentina}
\address{$^3$ Phyisics Department, Faculty of Science, University of Zagreb, Croatia}
\address{$^4$ Department of Physics, Texas A\&M University-Commerce,TX-USA}
\ead{krmpotic@fisica.unlp.edu.ar}

\begin{abstract}
 We extend the  formalism of weak interaction processes, obtaining
new  expressions for the transition rates,  which  greatly facilitate numerical calculations,
both for neutrino-nucleus reactions and muon capture.
We have done a thorough  study of exclusive (ground state)
properties of  $^{12}$B and $^{12}$N  within  the projected
quasiparticle random phase approximation (PQRPA). Good agreement
with experimental data is achieved  in this way.
The inclusive neutrino/antineutrino ($\nu/\tilde{\nu}$) reactions
$^{12}$C($\nu,e^-)^{12}$N and $^{12}$C($\tilde{\nu},e^+)^{12}$B
are calculated within both the PQRPA,  and the relativistic QRPA (RQRPA).
It is found that the magnitudes of the resulting cross-sections:
i) are  close to the sum-rule limit
at low energy, but significantly smaller than this limit
at high energies both for $\nu$ and $\tilde{\nu}$,
ii) they steadily increase when the  size of the configuration space is augmented,
and particulary for $\nu/\tilde{\nu}$ energies $> 200$ MeV,
and iii) converge  for sufficiently large configuration space and
final state spin.
\end{abstract}

\section{Introduction}

The neutrino-nucleus scattering on $^{12}$C
is important because this nucleus is a component of many
liquid scintillator detectors. As such it has been employed
in experimental facilities LSND, KARMEN, and LAMPF
to search for neutrino oscillations, and for measuring
neutrino-nucleus cross sections. The $^{12}$C target will be used  as well in
several planned experiments, such
as the spallation neutron source (SNS) at Oak Ridge National
Laboratory, and the Large Volume Detector (LVD)
at Gran Sasso National Laboratories. On the other hand, this nucleus is important
for astrophysics studies, as it forms one  of the onion-like shells of
 large stars before they
collapse.
Concomitantly, the LVD group have stressed recently
the importance of  measuring supernova neutrino oscillations.
 A comprehensive  theoretical review on neutrino-nucleus interaction was done by
 Kolbe \etal\cite{Kol03}.

\section{Formalism for the Weak Interacting Processes}

The most widely used formalism for neutrino-nucleus scattering
was developed by  the Walecka group~\cite{Wal95}. They classify
the nuclear transition  moments  as Coulomb,
longitudinal, transverse electric, and transverse magnetic.
This terminology is not
 used in  nuclear $\beta$-decay and  $\mu$-capture, where
one only speaks   on vector and axial matrix elements with different
degrees of   forbiddenness: allowed (GT and Fermi), first forbidden,
second forbidden, \etc.
Motivated by this fact,
we extend the  formalism of weak interaction processes developed in
\cite{Krm02,Krm05}, obtaining new  expressions for the transition rates,
which  greatly facilitate  numerical calculations, both for neutrino-nucleus
reactions and muon capture.

The weak Hamiltonian is expressed in the form~\cite{Wal95}
\br
H_{{\sss {W}}}(\rb)&=&\frac{G}{\sq}J_\alpha l_\alpha e^{-i\rb\cdot\kb},
\label{1}\er
where $G=(3.04545\pm 0.00006){\times} 10^{-12}$ is the Fermi coupling
constant (in natural units), the leptonic current $l_\alpha\equiv
\{ \lb,il_\emptyset\}$ is given by \cite[Eq. (2.3)]{Krm05} and
 the hadronic current operator $J_\a\equiv
\{ \Jb,i J_\emptyset \}$ in its nonrelativistic form \cite{Krm05}.
The quantity $k=P_i-P_f\equiv \{\kb,ik_\emptyset \}$
is the momentum transfer, ${\rm M}$ is the nucleon mass, and $P_i$ and $P_f$
are momenta of the initial and final nucleon
(nucleus). The effective  vector,
axial-vector, weak-magnetism and pseudoscalar
dimensionless coupling constants are defined in \cite{Krm05}.
In performing the multipole expansion of the nuclear operators
$O_\alpha\equiv (\Ob,iO_\emptyset)=J_\alpha e^{-i\kr}$
it is convenient: 1) to take $\kb$  along  the
$z$ axis, \ie  $e^{-i\kr}\go e^{-i\k z}$, and 2) to make use of
 Racah's algebra.  One obtains
\br
{O}_{\emptyset{\sf J}}&=&g_{\sss{V}}\M_{\sf J}^{\sss V}
+i\gas\M^{\sss A}_{\sf J}+i(\ga+\gpa)\M^{\sss A}_{0{\sf J}}
\label{12}\\
{O}_{{m}{\sf J}} &=&i(\delta_{{m}0}\gpb-\gA +m \gw)\M^{\sss A}_{{m}{\sf J}}
+\gvs\M^{\sss V}_{{m}{\sf J}}-\delta_{{m} 0}\gv\M_{\sf J}^{\sss V},
\label{13}\er
with the elementary operators
\br
&&\M^{\sss V}_{\sf J}=j_{\sf J}(\rho) Y_{{\sf J}}(\hat{\rb})
,\hspace{4cm}
\M^{\sss A}_{{m\sf J}}=\sum_{{\sf L}\ge 0}i^{ {\sf J-L}-1}
F_{{{\sf LJ}m}}j_{\sf L}(\rho)
\left[Y_{{\sf L}}(\hat{\rb})\otimes{\mbs}\right]_{{\sf J}},
\nn\\
&&\M^{\sss A}_{\sf J}=
{\rm M}^{-1}j_{\sf J}(\rho)Y_{\sf J}(\hat{\rb})(\mbs\cdot\mbn)
,\hspace{2cm}
\M^{\sss V}_{{m\sf J}}={\rm M}^{-1}\sum_{{\sf L}\ge 0}i^{ {\sf J-L}-1}
F_{{{\sf LJ}m}}j_{\sf L}(\rho)
[ Y_{\sf L}(\hat{\rb})\otimes\mbn]_{{\sf J}},
\nn\\
\label{14}\er
where $F_{{{\sf LJ}m}}$ are Clebsch-Gordan coefficients \cite{Krm05}.
The CVC  relates the vector-current pieces of
the operator $O_\alpha$ 
as 
$\kb\cdot\Ob^{\sss V}\equiv\k O_0^{\sss V}=
\tilde{k}_{\emptyset}O_\emptyset^{\sss V}$
with
$\tilde{k}_{\emptyset}\equiv k_{\emptyset}
-S(\Delta E_{\rm Coul}-\Delta\Mass)$,
where
$\Delta E_{\rm Coul}\cong \frac{6e^2Z}{5R} \cong 1.45 ZA^{-1/3}~~\mbox{MeV},$
is the Coulomb energy difference between the initial and final nuclei,
$\Delta\Mass=\Mass_n-\Mass_p=1.29$ MeV is the neutron-proton
mass difference, and $S=\pm 1$ for neutrino and antineutrino scattering, respectively.

The  transition amplitude
$\T_{{\sf J}^\pi_n}(\k)\equiv\sum_{ s_\ell,s_\nu }
\left|\bra{{\sf J}_n^\pi }H_{{\sss {W}}}(\k)\ket{0^+}\right|^{2}$
for the neutrino-nucleus reaction at a fixed value of $\k$,
from the initial state  $\ket{0^+}$ in the $(Z,N)$ nucleus to
the n-th final state $\ket{{\sf J}^\pi_n}$ in the
nucleus $(Z\pm 1,N\mp 1)$,  reads
\br
\T_{{\sf J}^\pi_n}(\k)
&=&{4\pi G^2}
[\sum_{\a=\emptyset,0,\pm 1} |\Bra{{\sf J}^\pi_n}{O}_{\a{ {\sf J}}}(\k)\Ket{0^+}|^2
\L_{\a}
-2\Re\left(\Bra{{\sf J}^\pi_n}{O}_{\emptyset{ {\sf J}}}(\k)
\Ket{0^+}
\Bra{{\sf J}^\pi_n}{O}_{0{ {\sf J}}}(\k)\Ket{0^+}^*\right)
\L_{\emptyset 0}],
\nn\\
\label{4}\er
where the momentum transfer is $k=p_\ell-q_\nu$, with
 $p_\ell\equiv\{\pb_\ell,iE_\ell\}$ and $q_\nu\equiv\{\qb_\nu,iE_{\nu}\}$.
The lepton traces
$\L_{\emptyset}, \L_{0}, \L_{\pm1}$ and $\L_{\emptyset 0}$
are  defined in \cite{Krm05}.

\section{ Neutrino (antineutrino)-nucleus cross section}

The exclusive  cross-section (ECS)
for the state $\ket{{\sf J}^{\pi}_n}$, as a function of the
incident neutrino energy $E_{\nu}$, is
\br
\s_\ell({\sf J}^{\pi}_n,E_{\nu})& = &\frac{|\pb_\ell|
E_\ell}{2\pi} F(Z+S,E_\ell)
\int_{-1}^1
d(\cos\theta)\T_{{\sf J}^\pi_n}(\kappa),
\label{25}\er
where $E_\ell=E_\nu-\w_{{\sf J}^\pi_n}$ and $|\pb_\ell|$ are the energy
and modulus of linear momentum of the lepton $\ell=e,\mu$,
and $\w_{{\sf J}^\pi_n}=-k_{\emptyset}=E_\nu-E_\ell$
is the excitation energy of the  state $\ket{{\sf J}^\pi_n }$ relative to
the  state $\ket{0^+}$. Moreover,  $F(Z+S,E_\ell)$ is
the Fermi function for neutrino $(S=1)$, and antineutrino $(S=-1)$  processes, respectively.
Here, we will also deal with the inclusive cross-sections (ICS),
\br
\s_\ell(E_{\nu})=\sum_{{\sf J}^{\pi}_n}\s_\ell({\sf J}^{\pi}_n,E_{\nu}),
\label{27}\er
as well as with folded  cross-sections, both exclusive, and inclusive
\br
\overline{\s}_\ell({{\sf J}^{\pi}_n})=\int dE_{\nu}\s_\ell({\sf J}^{\pi}_n,E_{\nu})
 n_\ell(E_{\nu})&,&
\overline{\s}_\ell= \int dE_{\nu}
\s_\ell(E_\nu) n_\ell(E_{\nu}),
\label{28}\er
where $n_\ell(E_{\nu})$ is the neutrino~(antineutrino) normalized flux.
In  the evaluation of both neutrino, and antineutrino ICS the summation
in \rf{27} goes over all $n$ states with spin and parity
${\sf J}^{\pi}\le 7^{\pm}$  in the PQRPA (Projected Quasiparticle Random Phase
Approximation), and
over ${\sf J}^{\pi}\le 14^{\pm}$ in the RQRPA (Relativistic QRPA).

\section{Numerical results}

\begin{figure}[h]
\centering
\begin{tabular}{cc}
\hspace{-1cm}
{\includegraphics[width=9 cm,height=9 cm]{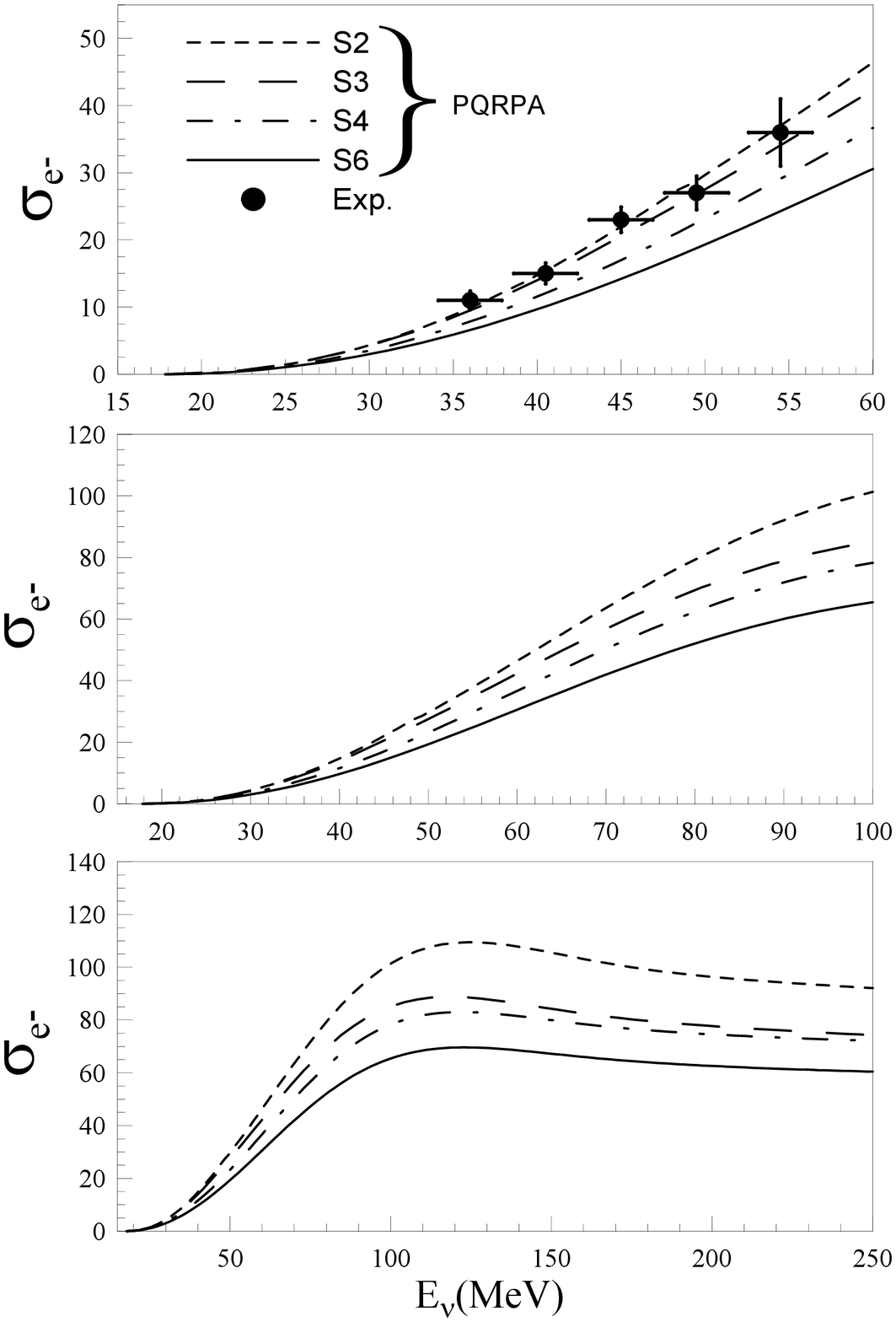}}
\vspace{-.25cm}
&\hspace{-2cm}
{\includegraphics[width=9 cm,height=9 cm]{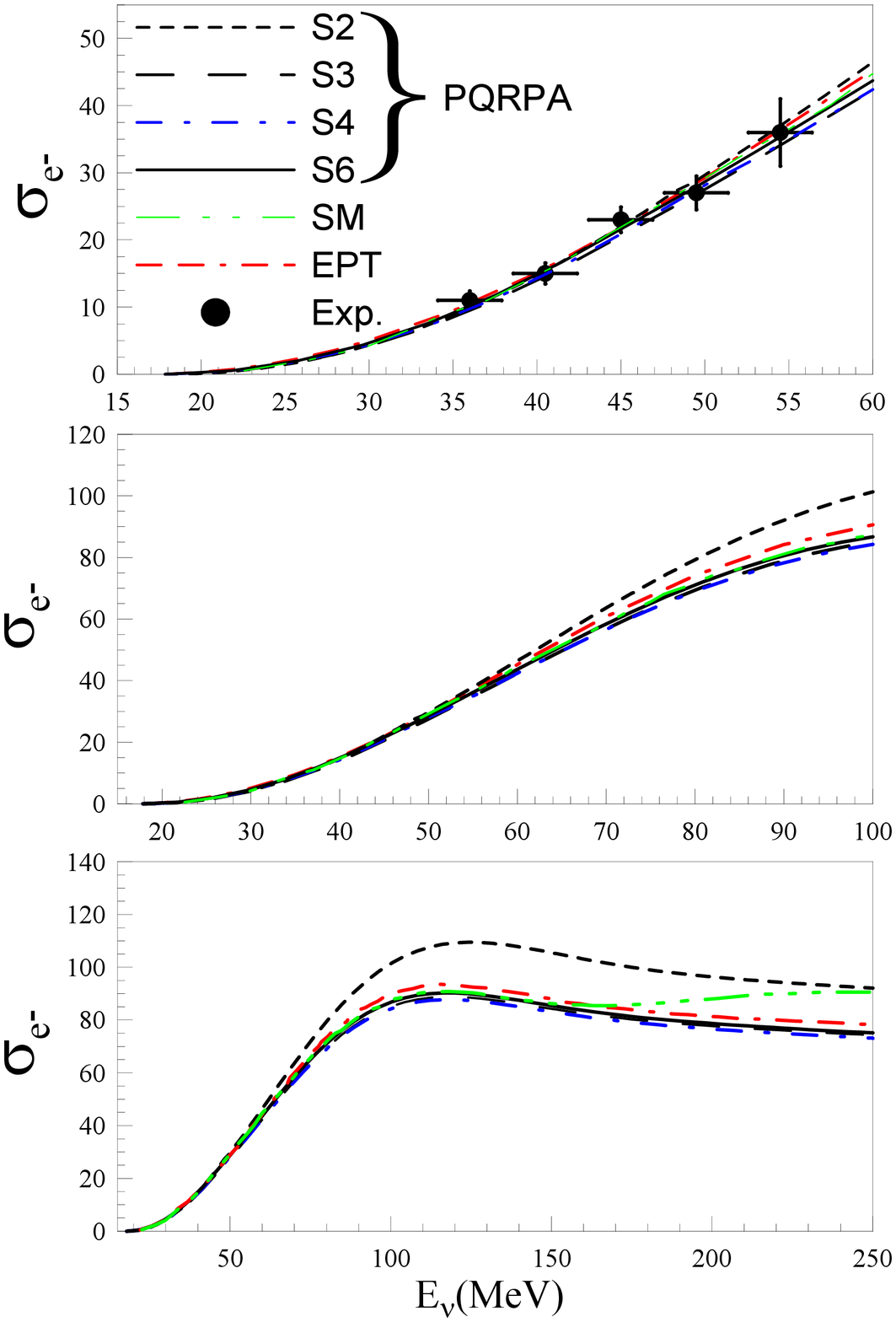}}\vspace{-.25cm}
\end{tabular}
\caption{\label{F1}(Color online) ECS $\s_{e^-}(1^+_1,E_{\nu})$ for
the reaction $^{12}$C($\nu_e,e^-)^{12}$N (in units of
$10^{-42}$ cm$^2$), as a function of the incident neutrino
energy $E_\nu$. On the left side  we have $t=0$ for all $S_N$,
whereas for the right side $t=0$ for $S_2$, and  $S_3$, $t=0.2$
for $S_4$, and $t=0.3$ for $S_6$. The experimental data in  the
DAR region are from Ref.~\cite{Ath97}.}
\end{figure}
 The calculations were performed with the PQRPA using the model spaces
$S_2, S_3, S_4$,  and $S_6$ including 2, 3, 4, and 6 $\hbar
\omega$ harmonic oscillators shells, respectively, and by
employing a $\delta$-interaction.
 For $S_2$, $S_3$, and $S_4$ spaces the  s.p. energies and pairing
strengths  were varied in a $\chi^2$ search to account
for the experimental spectra of odd-mass nuclei $^{11}$C, $^{11}$B, $^{13}$C, and $^{13}$N
~\cite{Krm05}. As this method can not be used  for the  $S_6$ space,
which comprises $21$ s.p. levels, the energies in this case were derived
as in  Ref.~\cite{Paa07}, while the pairing strengths
were adjusted to reproduce the experimental
gaps in $^{12}$C~\cite{Sam09}.
We also employ the RQRPA theoretical framework ~\cite{PNVR.04} with  $S_{20}$, and $S_{20}$  spaces.
In this case, the ground state is calculated in the Relativistic
Hartree-Bogoliubov model using effective Lagrangians with density
dependent meson-nucleon couplings and DD-ME2 parameterization, and
pairing correlations are described by the finite range Gogny force.

Fig.  \ref{F1} shows  ECS $^{12}$C$_{gs}$($\nu_e,e^-)^{12}$N$_{gs}$
plotted as a function of the incident neutrino energy $E_\nu$ for
different final energy region $E_\nu\leq 60$ MeV (DAR region),
$E_\nu\leq 100$ MeV ( supernovae neutrino signal search) and
$E_\nu\leq 250$ MeV ( neutrinos oscillation search in LSND).
On  the left side the
$\delta$-interaction particle-particle strength is $t=0$ for all $S_N$
\footnote{This value for $S_3$ was adopted in Ref.~\cite{Krm05}.}, whereas on
the right side $t$ is gauged  to reproduce the ground state energy of $^{12}$N and the $B(GT)$ values of $^{12}$N
and $^{12}$B,
getting $t=0$ for $S_2$ and  $S_3$, $t=0.2$ for $S_4$, and $t=0.3$ for $S_6$.
The ECS experimental data in  the DAR region are from Ref.~\cite{Ath97}.
\begin{figure}
\centering
\begin{tabular}{cc}
\hspace{-1cm}
{\includegraphics[width=8 cm,height=9.1 cm]{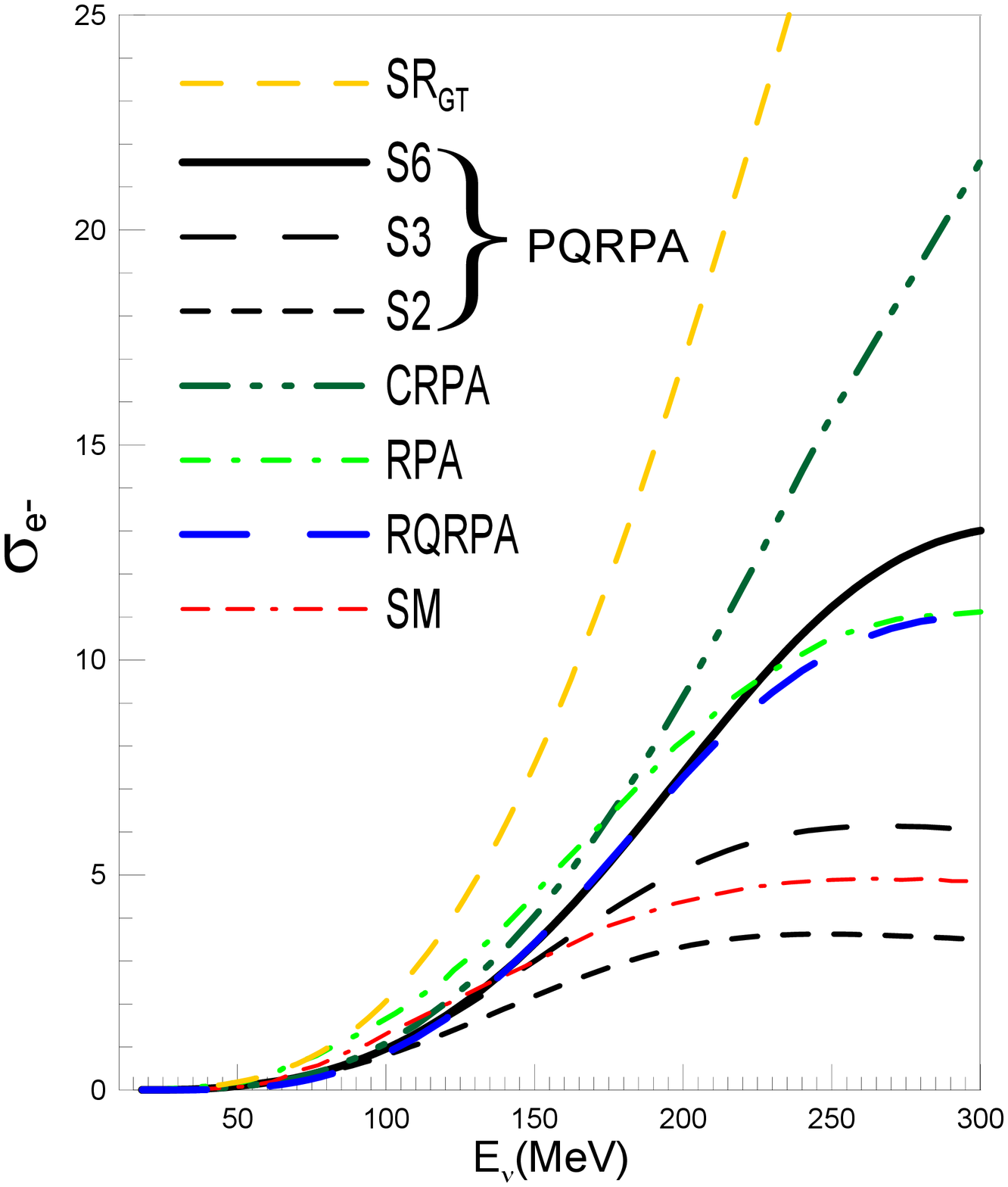}}\vspace{-.5cm}
&\hspace{-1cm}
{\includegraphics[width=8 cm,height=9 cm]{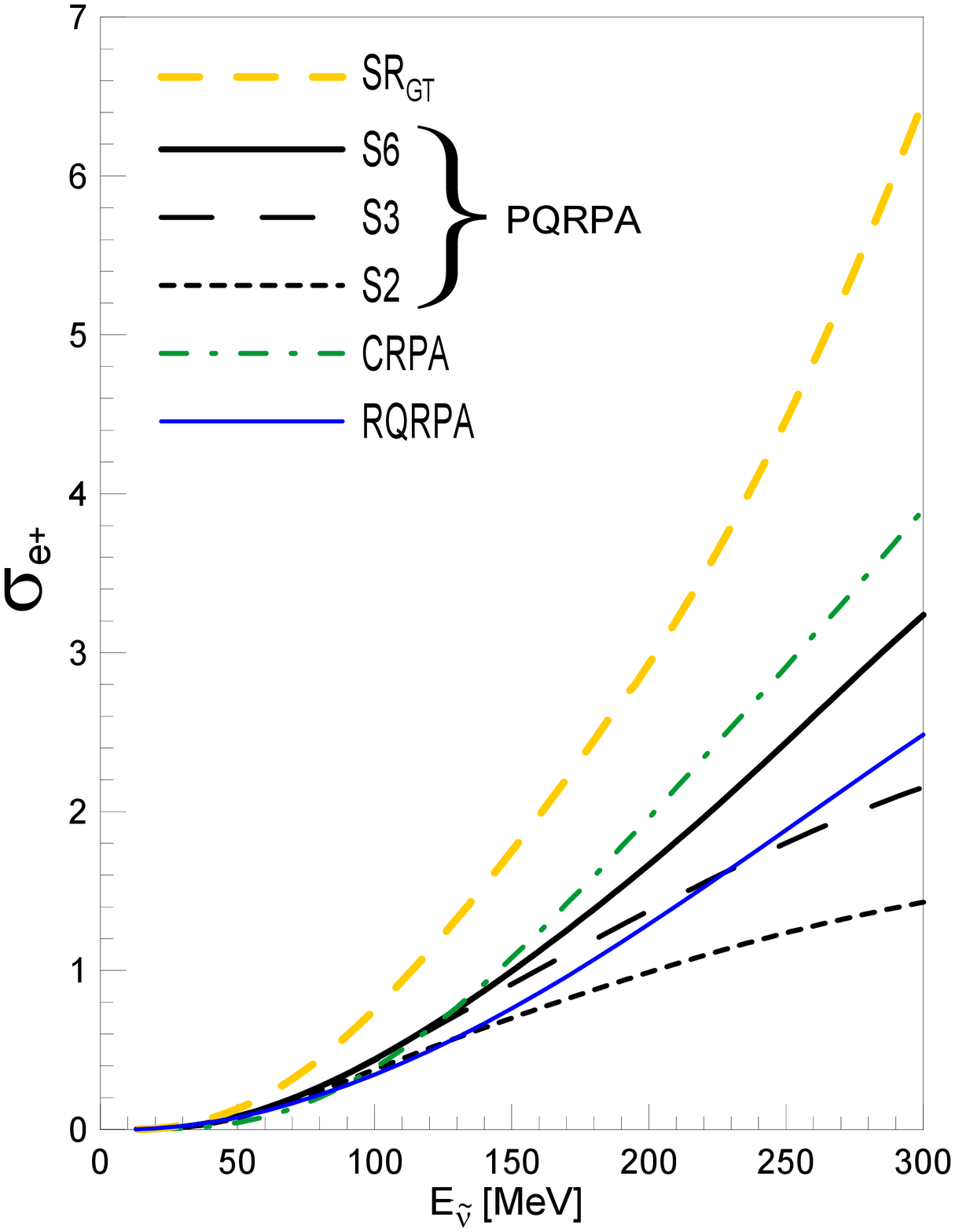}}\vspace{-.5cm}
\end{tabular}
\caption{\label{F2}(Color online)
Inclusive cross sections  $\sigma_{e^-}(E_\nu)$, and $\sigma_{e^{+}}(E_{\tilde\nu})$ (in
units of $10^{-39}$ cm$^2$) for the reactions
 $^{12}$C($\nu_e,e^{-})^{12}$N, and
 $^{12}$C(${\tilde \nu}_e,e^{+})^{12}$B, respectively,
plotted as a function of incident
neutrino and antineutrino energies. The PQRPA results within the s.p. spaces
$S_2$, $S_3$, and $S_6$, have the same parametrization  as those on the
right panel  of Fig.~\ref{F1}. The sum rule limit $SR_{GT}$ and several
previous RPA-like calculations, namely: RPA~\cite{Vol00}, CRPA~\cite{Kol99b}, and
RQRPA within  $S_{20}$ for  two-quasipartile cutoff $E_{2qp}$=100 MeV~\cite{Paa07}, as well
as the shell model result
~\cite{Vol00}, are also exhibited.}
\end{figure}

On the left and right panels of  Fig.~\ref{F2} are displayed  the
 $^{12}$C($\nu_e,e^-)^{12}$N and $^{12}$C(${\tilde\nu}_e,e^+)^{12}$B ICS  $\sigma_{e^{\mp}}(E_\nu)$ , respectively.
The PQRPA results within the s.p. spaces $S_2$, $S_3$ and $S_6$,
have the same values of $t$ as in the right panel of Fig.~\ref{F1}.
They are compared with: i) the sum rule limit $SR_{GT}$ obtained with
an average excitation energy $\overline{\w_{{\sf J}^\pi_n}}$ of $17.34$ MeV, ii)
 other RPA-like calculations, and  iii) the shell model (SM).

\begin{figure}[h]
\centering
\begin{tabular}{cc}
\hspace{-1.cm}
{\includegraphics[width=8.6 cm,height=10 cm]{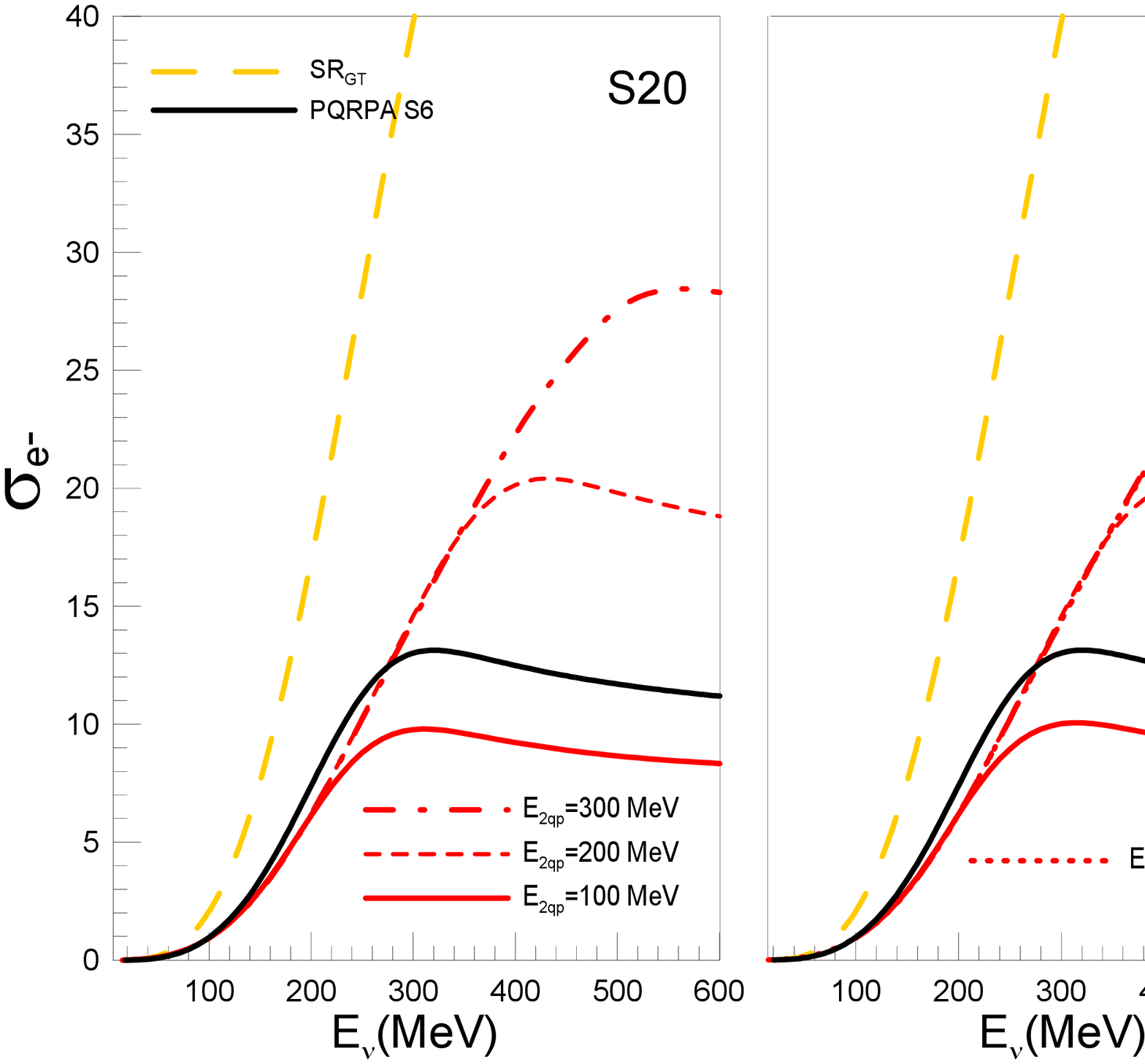}}\vspace{-1.5 cm}
&\hspace{-.5cm}
{\includegraphics[width=8.6 cm,height=10 cm]{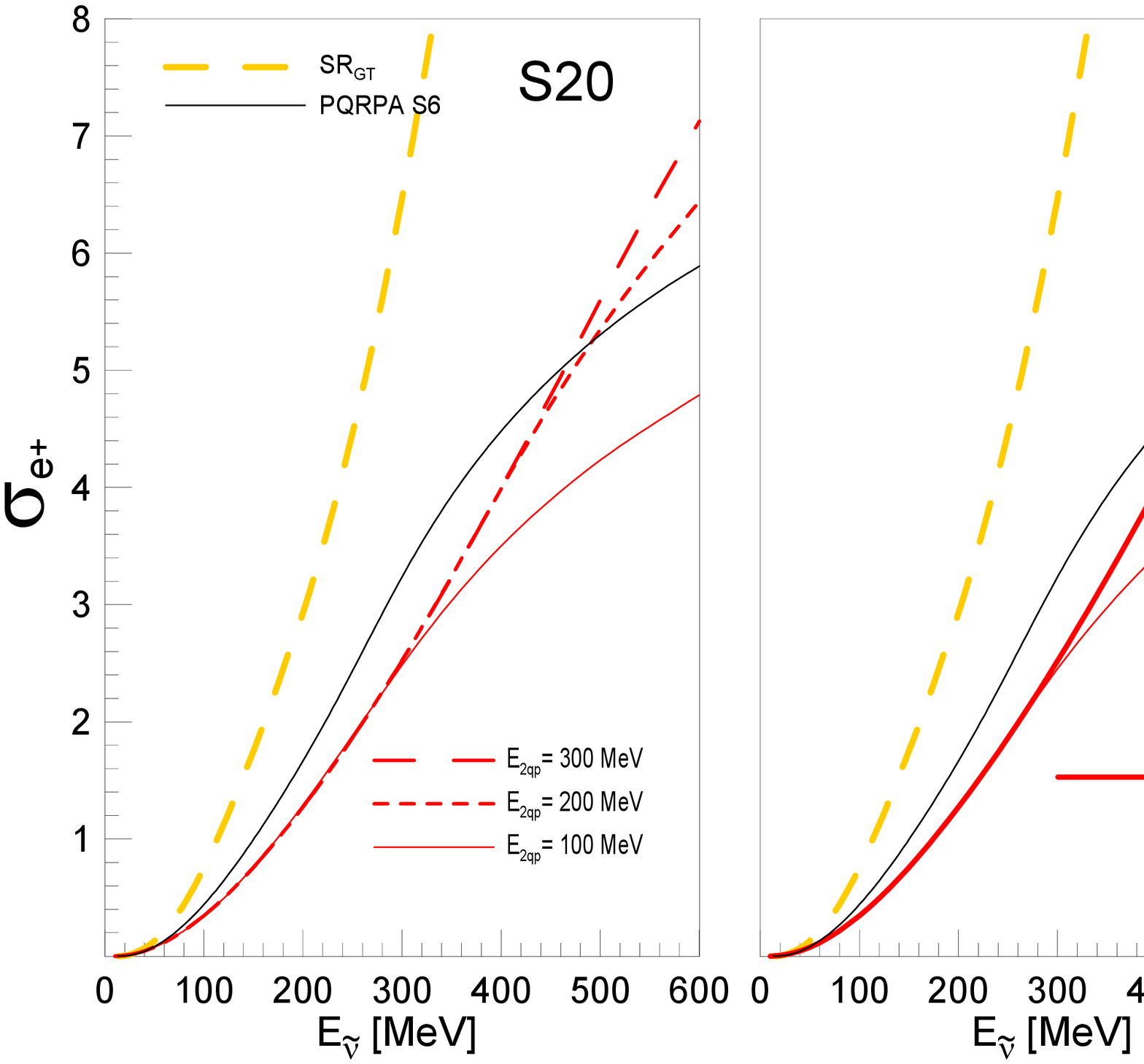}}\vspace{-1.5 cm}
\end{tabular}
\caption{\label{fig3}(Color online) Inclusive cross sections
$^{12}$C($\nu_e,e^-)^{12}$N (left panel) and $^{12}$C(${\tilde
\nu}_e,e^+)^{12}$B (right panel) (in units of $10^{-39}$ cm$^2$)
 evaluated within  the RQRPA for different
configuration spaces.}
\end{figure}

 In Fig.~\ref{fig3} are plotted the $\sigma_{e^{\mp}}(E_\nu)$, evaluated in RQRPA with different
configuration spaces according the  values of  quasiparticle energy cut-off   $E_{qp}$.
One sees that  the ICS increase as  the configuration space  increase, saturating
 at higher and higher energies.

\begin{figure}
\centering
{\includegraphics[width=14 cm,height=9 cm]{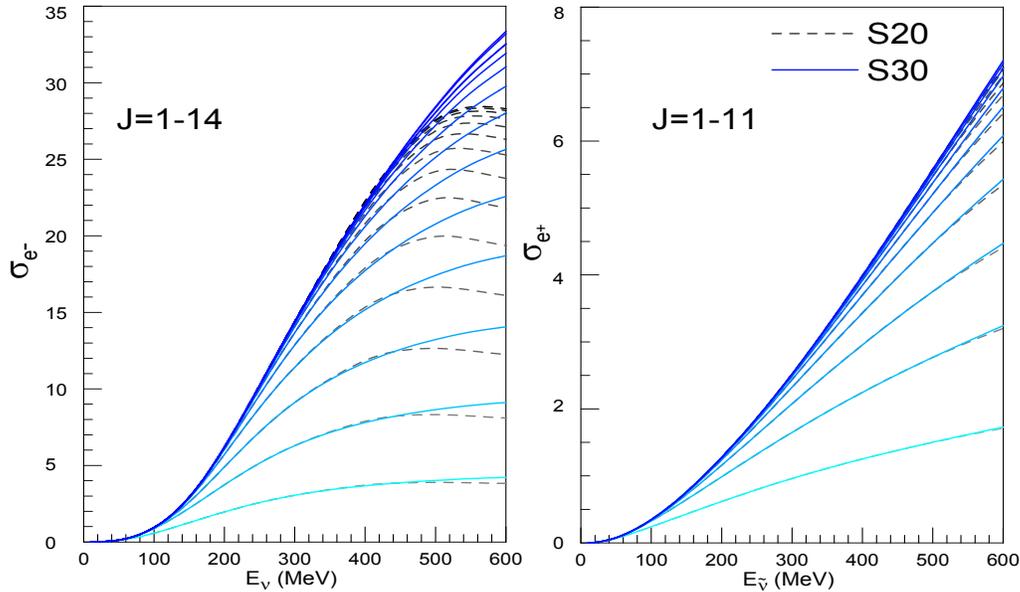}}\vspace{-3. cm}
 \caption{\label{fig4}(Color
online) Inclusive cross sections $^{12}$C($\nu_e,e^-)^{12}$N (left panel) and
$^{12}$C(${\tilde\nu}_e,e^+)^{12}$B (right panel) (in units of
$10^{-39}$ cm$^2$) evaluated within  the RQRPA for $E_{2qp}=500$ MeV and spaces $S_{20}$,
and $S_{30}$,  for different maximal spins, which  go up to $J=14$ for neutrinos,
and up to $J=11$ for
antineutrinos. }
\end{figure}

Finally, Fig.~\ref{fig4} presents the  RQRPA results for
$\sigma_{e^{\mp}}(E_\nu)$ with different maximal spins $J$. One
sees that while for antineutrinos contribute  spins only up to
$J=11$, it is needed  to go up to $J=14$ in the case of neutrinos.
It is also clear that to achieve the saturation one should go to
energies larger than $600$ MeV. ~\bigskip ~\vspace{2cm}
\newpage
~\bigskip
~\vspace{1.cm}
~\vspace{-1cm}

\newpage

\vspace{-0.5cm}
\section{Summary}
The shell model and  the PQRPA are proper theoretical frameworks to describe the
ground state properties of $^{12}$B and  $^{12}$N and the  ECS $\s_{e^-}(1^+_1,E_{\nu})$ for
the reaction $^{12}$C($\nu_e,e^-)^{12}$N.
The ICS calculated,  at difference with the exclusive ones,
steadily increase, and particulary for neutrino energies larger
than $200$ MeV, when the  size of the configuration space is
augmented, in spite of including the particle-particle interaction.
They approach to those of the first-forbidden sum-rule limit at
low energy, but are significantly smaller at high energies
both for neutrino and antineutrino.
The study of the partial ICS's has been related with the proposal done in Ref.~\cite{Laz07}
of performing nuclear structure studies of forbidden processes by using low energy
neutrino and antineutrino beams.
We conclude  that to study forbidden
reactions in $^{12}{\rm C}({\tilde{\nu}},e^+)^{12}{\rm B}$ process,
one would need  ${\tilde{\nu}}$-fluxes with
$E_{\tilde{\nu}}$ up to  $\geq 150$ MeV, 
while  those from Ref.~\cite{Laz07} go up  to $80$ MeV only
due to the feasibility of the proposed experiments.

\section*{Acknowledgements}
This work was partially supported by the Argentinean agency CONICET under
contract PIP 0377, and by the U.S. DOE grants
DE-FG02-08ER41533, DE-FC02-07ER41457 (UNEDF, SciDAC-2) and the Research Corporation.
N. P. acknowledges support by the Unity through Knowledge Fund
(UKF Grant No. 17/08),  MZOS - project 1191005-1010
 and Croatian National Foundation for Science.

\end{document}